\documentstyle[12pt]{article}
\begin{document}

\rightline{FTUV 94-64 / IFIC 94-61}
\rightline{Oct. 1994, rev. 28 Feb. 1995}
\rightline{hep-th 9411121}

\begin{center}
{\Large{\bf On the physical contents of q-deformed Minkowski spaces}}
\end{center}

\vspace{1\baselineskip}

\begin{center}
{\large{J.A. de Azc\'{a}rraga$^{\dagger}$, P.P. Kulish\footnote{On
leave of absence from the St.Petersburg's Branch of the Steklov Mathematical
Institute of the Russian Academy of Sciences.}$^{\dagger}$
and F. Rodenas$^{\dagger \; \ddagger}$}}
\end{center}

\addtocounter{footnote}{-1}

\noindent
{\small{\it $ \dagger$ Departamento de F\'{\i}sica Te\'{o}rica and IFIC,
Centro Mixto Universidad de Valencia-CSIC
E-46100-Burjassot (Valencia), Spain.}}

\noindent
{\small{\it $ \ddagger $ Departamento de Matem\'atica Aplicada,
 Universidad Polit\'ecnica de Valencia
E-46071 Valencia, Spain.}}

\vspace{1\baselineskip}

\begin{abstract}
Some physical aspects of $q$-deformed   spacetimes are
discussed. It is pointed out that, under certain standard assumptions
relating deformation and quantization, the classical
limit (Poisson
bracket  description) of the dynamics
is bound to contain unusual features. At the same time, it is argued
that the formulation of an associated $q$-deformed field theory is fraught
with serious difficulties.
\end{abstract}

\section{Introduction}

Since  the early days of quantum theory there have been attempts to
quantize the spacetime manifold, the latest one being probably related
to superstring theory. This question is now being actively  discussed
using different arguments and from different backgrounds (see, {\it e.g.},
\cite{DFR} and references therein).

A recent  generalization of  Lie groups and algebras, the quantum groups
(or deformations of the algebra of functions on Lie groups and of their
enveloping algebras) provides another framework
to construct non-commutative spacetime coordinates. This is achieved
by deforming the
kinematical  Lie groups and in particular the Poincar\'e and Lorentz groups
\cite{WSSW,OSWZ}. In this scheme, a quantum Minkowski spacetime ${\cal M}_q$
may
be  introduced by extending to the quantum  ($q \neq 1$) case the `classical'
or undeformed ($q=1$)  Lie group construction,
by which spacetime (a four-vector)  is constructed out of two (dotted and
undotted) spinors.

It has been pointed out \cite{AKR} that the $R$-matrix \cite{FRT} and
the reflection
equations  (RE) (see  \cite{K-SK,K-SA} and references therein as well as
\cite{MBRA,MAJID,MEYER}
in the context of braided quantum groups) constitute a suitable formalism
to describe in an unified manner
certain $q$-deformations of the Minkowski spacetime. The physical contents
of these deformations is, however, up to now  unclear. This letter is
an attempt to
discuss some general features of these deformations to see whether there are
grounds to favour some of them and to check the consistency of the
interpretations. In fact,  some mathematical
\cite{WSSW,OSWZ,MAJID,MEYER,LRTN2,K,PSW}
and physical (see, in particular,
\cite{LRR,BACRY,BMT}) properties of deformed Poincar\'e algebras
have been recently discussed  but it is fair to say that a complete
picture of a $q$-deformed physical relativistic system (not to say of a
quantum field theory) is still missing.

Such a picture should include:

\vspace{0.5\baselineskip}
\noindent
1. a suitable spacetime non-commutative coordinate algebra ${\cal M}_q$

\noindent
2. a deformed Poincar\'e group and its coaction on the spacetime
coordinates

\noindent
3. a deformed Poincar\'e algebra, part of it generated by the deformed
momenta

\noindent
4. an appropriate  definition of  phase space from the $q$-spacetime
coordinates and
$q$-momenta, as well as the associated algebra of observables.

\vspace{0.5\baselineskip}

 Apart from these considerations, a precise relation between $q$-deformation
($q$$ \neq $1) and quantization ($ \hbar \neq 0$) has to be postulated
if $q$ and $\hbar$ are not independent constants (see below).
In fact, once  commutation relations among the $q$-coordinates and
$q$-momenta covariant under the action of the $q$-Poincar\'e group
are determined, it is possible to interpret them as:

\vspace{0.5\baselineskip}
\noindent
a) the algebra of quantum observables, and then study its
irreducible representations (see \cite{PSW,K,AKR2})

\noindent
b) as a `classical' algebra of `$q$-numbers' with a possible further
quantization to introduce Planck's constant (here, usually a $q$-path
integral formalism is invoked, see {\it e.g.} \cite{BFC}).

\vspace{0.5\baselineskip}
The second interpretation is in line with the quantization of mechanics with
Grassmann variables and supersymmetry. Here we shall take the
`standard'  point of view, so that non-commuting quantities will
already
be considered as operators. As a result, the elements of the deformed
Poincar\'e group(s) must be treated also as operators. This confers the group
parameters a certain dynamical character   absent in the undeformed case,
in which they are real numbers. Thus, the symmetry and the transformation
properties
of a relativistic system with respect to such a quantum group are different
from  those studied in \cite{MS,RS}, where the ground state and
the Hamiltonian  are invariant under the coaction of the quantum group, which
plays
the r\^ole of an internal symmetry and is not the quantum kinematical group
itself.

Since the one-particle case already provides the basic ground to analyze
the above problems, we shall not consider here the question of
multiparticle systems, which lead to the braided Hopf algebra
structure of quantum Minkowski spacetime \cite{MBRA,MAJID,MEYER,MARU}.

\section{Possible $q$-Minkowski spaces}

\indent
To introduce non-commutative `coordinates' for spacetime it is natural
to assume that this non-commutativity is determined  by their transformation
properties under the corresponding quantum Lorentz group.
The definitions for this group  to be considered here will
all be based on the well established
$SL_q(2,C)$ quantum group. This is  defined
through an `RTT' relation \cite{FRT} for the 2$\times$2 matrix $M$ of
the $SL_q(2,C)$ generator
\begin{equation}\label{2a}
R_{12}M_1M_2=M_2M_1R_{12} \quad,
\end{equation}
\noindent
where $M_1=M \otimes 1$, $M_2=1 \otimes M$ and $R_{12}$ is the $SL_q(2,C)$
4$\times$4 $R$-matrix.

Thinking of the classical homomorphism $SL(2,C) \rightarrow L$, the spacetime
`coordinates' are defined as the entries of a 2$\times$2 matrix $K$ (the
analog of $ \sigma_{\mu}x^{\mu}$)  transforming as
\begin{equation}\label{2b}
K \longmapsto K'= MK M^{\dagger} \quad,\quad K=K^{\dagger} \quad ,
\end{equation}
\noindent
where in the quantum case  $ (M^{\dagger})^{-1}$
is an independent copy of the $SL_q(2,C)$ generators matrix
also satisfying (\ref{2a}),
$R_{12}(M_1^{\dagger})^{-1}(M_2^{\dagger})^{-1}
= (M_2^{\dagger})^{-1}(M_1^{\dagger})^{-1}R_{12}$ with $q$ real and
${\cal P}R_{12}{\cal P}=R_{21}=R_{12}^{\dagger}$
(${\cal P}$ is the permutation  operator in $C^2 \otimes C^2$). The
commutation properties of the generators of a  $q$-Minkowski algebra
defined by means of $K$ may be expressed by a RE  as \cite{AKR}
\begin{equation}\label{2c}
R^{(1)}K_1R^{(2)}K_2=K_2R^{(3)}K_1R^{(4)} \quad.
\end{equation}
\noindent
The covariance condition now defining the $q$-Minkowski algebra is expressed
by saying that the coaction (\ref{2b}) preserves (\ref{2c}). This gives
\begin{equation}\label{2d}
R^{(1)}M_1M_2=M_2M_1R^{(1)} \quad, \qquad
M^{\dagger}_1R^{(2)}M_2 =M_2R^{(2)}M_1^{\dagger}
\end{equation}
\noindent
plus $R^{(1)}=R^{(4)\,\dagger}$ and $R^{(3)}= {\cal P}R^{(2)}{\cal P}^{-1}$
($R^{(3)}_{12}=R^{(2)}_{21}$). Thus, the coaction (\ref{2b}) plus the
preservation of the algebra (\ref{2c}) does not lead to a unique
$q$-Minkowski algebra ${\cal M}_q$  or $q$-Lorentz group (\ref{2d}),
since there are many consistent
solutions for the 4$\times$4 matrices $R^{(i)}$ ($i=1,...,4$) in (\ref{2c}).
Let us point out some of them

\vspace{0.5\baselineskip}
\noindent
1) \,  $R^{(1)}=R_{12}$, $R^{(2)}=R_{21}$. This defines the $q$-Minkowski
space ${\cal M}_q^{(1)}$ \cite{WSSW,OSWZ}.

\noindent
2) \, $R^{(1)}=R_{12}$, $R^{(2)}=I_4$. This possibility leads
to a  ${\cal M}_q^{(2)}$   algebra \cite{WSSW,AKR2,MAJ}
isomorphic to the $GL_q(2)$ quantum group.

\noindent
3) \, $R^{(1)}=I_4$, $R^{(2)}=V \equiv diag(q^2,1,1,q^2)$.
This defines  a `twisted' Minkowski space ${\cal M}_q^{(3)}$ (we shall
use $q$ although this case \cite{CHA-DEM}  is not a proper deformation).

\vspace{0.5\baselineskip}
\noindent
Once we have commutation properties preserved under a certain Lorentz
quantum group
defined by (\ref{2d}) for a specific choice of $R^{(i)}$, the corresponding
$q$-Lorentz algebra which acts on coordinates may be introduced by
duality bet\-ween them as dual Hopf algebras \cite{FRT,DRINFELD}.
Also, the $q$-momenta to be introduced below (the translation part of
the $q$-Poincar\'e algebra) may be related to $q$-coordinates by duality in a
general setting \cite{MARU}. However, as mentioned in the introduction
we shall restrict ourselves to the algebraic aspects of ${\cal M}^{(i)}_q$.

The translation part of a $q$-Poincar\'e algebra ($q$-momenta) is
introduced in this scheme by means of the non-commuting entries of
a 2$\times$2  matrix $Y$ which satisfies
\begin{equation}\label{2e}
R^{(1)}Y_1R^{(3)\, -1}Y_2=Y_2R^{(2)\, -1}Y_1R^{(4)} \quad.
\end{equation}
\noindent
The matrices $R^{(i)}$ for which (\ref{2e}) is invariant under the
coaction
\begin{equation}\label{2e1}
Y \longmapsto Y'=(M^{\dagger})^{-1} YM^{-1} \quad,
\end{equation}
\noindent
determine through (\ref{2e}) the
corresponding $q$-derivative algebras ${\cal D}_q^{(i)}$. Because of the
transformation  properties (\ref{2b}) and (\ref{2e1}),
we may call $Y$ `covariant' if $K$ is `contravariant'. The
$q$-Lorentz invariant scalar product is given \cite{AKR} by the $q$-trace
\cite{FRT,ZUM}. For instance, a $q$-analogue of the dilatation operator
$s=x^{\mu} \partial_{\mu}$ is given by
\begin{equation}\label{2f}
s= tr_q(KY)=tr ({\cal D}KY) \quad,
\end{equation}
\noindent
where ${\cal D} \propto tr_{(2)}({\cal P}[(R^{(1)\,t_1})^{-1}]^{t_1})$
(the trace is in the second space and the transposition in the first one;
the proportionality factor  is fixed by convenience being 1 for $q$=1).
Mixed commutation relations are defined by an invariant inhomogeneous RE
\cite{AKR}
\begin{equation}\label{2g}
Y_2 R^{(1)}K_1R^{(2)}=R^{(3)}K_1R^{(1)\,-1}Y_2 + J  \quad
\end{equation}
\noindent
which extends to the $q$-case the familiar relation $\partial_{j}x^{i}=
x^{i} \partial_{j} + \delta^i_j$. A complete $q$-differential
calculus (see \cite{OSWZ} for ${\cal M}_q^{(1)}$) may be developed within this
scheme \cite{AKR,AKR2}.

In order to extract some physical consequences of the $q$-deformation it
is natural to consider   the simplest ${\cal M}^{(3)}_q$ case
\cite{CHA-DEM} since, as it has been pointed out \cite{LRTN}, it
corresponds to a twisted algebra \cite{DRINFELD,RE} situation
({\it i.e.}, not a proper
$q$-deformation) and thus it must be simpler as the diagonal form of $V$
above already suggests. The defining properties of the
coordinate and derivative algebras (\ref{2c}), (\ref{2e}), (\ref{2g})
may be expressed as
\begin{equation}\label{2h}
\begin{array}{c}
Z_1VZ_2=Z_2VZ_1 \quad, \qquad D_1V^{-1}D_2=D_2V^{-1}D_1 \quad , \\
\, \\
D_1Z_2= VZ_2D_1V^{-1} + {\cal P} \quad .
\end{array}
\end{equation}
\noindent
where we have relabelled $K$=$K^{(3)}$=$Z$ in (\ref{2c}), and in
(\ref{2e}), $Y$=$Y^{(3)}$=$D$. Defining
\begin{equation}\label{2h1}
Z= \left( \begin{array}{cc}
           z^1 & z^4 \\
           z^2 & z^3
           \end{array} \right) \; , \; Z'=MZM^{\dagger} \;,\;
D=  \left( \begin{array}{cc}
           \delta_1 & \delta_2 \\
          \delta_4 & \delta_3
           \end{array} \right) \; ,\;
D'=(M^{\dagger})^{-1}DM^{-1} \;
\end{equation}
\noindent
($Z$=$Z^{\dagger}$, $D$=$-D^{\dagger}$; we denote by $\delta_i$
the non-commuting derivatives and reserve
$\partial_i$ for the ordinary commuting ones)
the commuting properties of the $z^i$, $\delta_i$ are easily found from
eqs. (\ref{2h}).
As an example, the first  eq. in (\ref{2h}) gives {\it e.g.} \cite{CHA-DEM}
\begin{equation}\label{zz}
\begin{array}{lll}
z^1z^2=q^2z^2z^1 \quad,  & \quad z^1z^3=z^3z^1 \quad, & \quad z^4z^1=q^2z^1z^4
\quad, \\
z^2z^3=q^2z^3z^2 \quad , & \quad z^2z^4=z^4z^2 \quad, & \quad
z^3z^4=q^2z^4z^3 \quad.
\end{array}
\end{equation}

Given the matrix elements of an algebra of one  type, contravariant or
covariant
(cf. eqs. (\ref{2b}), (\ref{2e1})),
we may use the 4$\times$4  matrix
$\hat{V}^{\epsilon}= \epsilon_2 {\cal P}V \epsilon_2$ ($\epsilon_2= 1 \otimes
\epsilon$,
$\epsilon = i \sigma_2$,
$\epsilon M \epsilon^{-1}= (M^{-1})^t$),
\begin{equation}\label{2l}
\hat{V}^{\epsilon}= \left( \begin{array}{cccc}
                            0 & 0 & 0 & 1 \\
                            0& -q^2 & 0 & 0 \\
                            0 & 0& -q^2 & 0 \\
                            1 & 0 & 0 & 0
                           \end{array} \right) \quad , \quad
\end{equation}
\noindent
to obtain coordinates of the other type since
\begin{equation}\label{2l1}
\hat{V}^{\epsilon}(M \otimes M^* ) \hat{V}^{\epsilon \, -1} =
M^{-1 \, \dagger} \otimes M^{-1 \, t} \quad .
\end{equation}
\noindent
This change of type will be denoted by adding (or removing) an
overbar to the original matrix. For instance,
$ \bar{Z}=\hat{V}^{\epsilon}Z$,
$\bar{Z}' =(M^{\dagger})^{-1} \bar{Z} M^{-1}$, is covariant
(if we write $\bar{Z}_{ij}=
\hat{V}^{\epsilon}_{ij,kl}Z_{kl}$, $i,j=1,2$, the vector $Z_{kl}$ is
$(z^1,z^4,z^2,z^3)$ and $\bar{Z}_{ij}$ is $(z_1,z_2,z_4,z_3)$).
Similarly, $ \bar{D}=\hat{V}^{\epsilon\,-1}D$
is contravariant ($\bar{D}' =M \bar{D} M^{\dagger}$)
because of (\ref{2l1})).
As a result, $tr(\bar{Z}Z)=tr(Z\bar{Z})$ is an invariant and defines the
scalar product
\begin{equation}\label{2m}
(Z,Z) \equiv  \frac{1}{2} tr( \bar{Z}Z)= z^1z^3-q^2z^2z^4 \qquad
( \, \bar{Z}= \left( \begin{array}{cc}
                  z^3 & -q^2 z^4 \\
                 -q^2 z^2 & z^1
                \end{array} \right) \,) \;.
\end{equation}
\noindent
In fact, $(Z,Z)=\frac{1}{2} Z_{ij}g_{ij,kl} Z_{kl}$ defines
the $q$-Lorentz invariant metric for ${\cal M}_q^{(3)}$,
\begin{equation}\label{nmetric}
g=\hat{V}^{\epsilon } {\cal P}= \left( \begin{array}{cccc}
                               0&0&0&1 \\
                               0&0&-q^{-2}&0 \\
                               0&-q^{-2}&0&0 \\
                               1&0&0&0
                              \end{array} \right) \quad.
\end{equation}
\noindent
If we write $(Z,Z)= \frac{1}{2} g_{ij}z^iz^j$ with $i,j=1,2,3,4$, then
$g_{13}=g_{31}=1$ and  $g_{24}=g_{42}=-q^{-2}$.
A similar procedure \cite{AKR2}
may be used to introduce the metric in the other cases; for the present
${\cal M}_q^{(3)}$ case  the trace in (\ref{2m}) is
the ordinary trace, cf. (\ref{2f}).

An important consequence of the simple (twisted) nature of the
(diagonal) $R$-matrices defining ${\cal M}_q^{(3)}$
is that its  non-commuting properties may be accounted for
by introducing two operators $u$, $v$ such that
\begin{equation}\label{2i}
vu=q^2uv \quad, \quad u^{-1}= u^{\dagger} \;,\; v=v^{\dagger} \;.
\end{equation}
\noindent
If we now define $Z$ in (\ref{2h1}) as
\begin{equation}\label{2j}
Z \equiv \left( \begin{array}{cc}
          v x^1 & u^{-1}x^4 \\
           ux^2  & v^{-1}x^3
           \end{array} \right) \quad, \quad X= \left( \begin{array}{cc}
           x^1 & qx^4 \\
          qx^2 & x^3
           \end{array} \right)
\end{equation}
\noindent
it is simple to see that the commutation properties (\ref{2i}) and
$Z_1VZ_2=Z_2VZ_1$, imply that the $x$ components are commuting,
$X_1X_2=X_2X_1$.
This reduction to the commutative case also applies to the
commutation relations that two vectors $Z$ and $Z'$ have to
satisfy so that their linear combination $\alpha Z + \beta Z'$ is isomorphic
to $Z$ (cf. \cite{CHA-DEM}) {\it i.e.}, that the braided coaddition (cf.
\cite{MEYER}) is defined. In the present case,
the braiding is given by
$Z_1VZ'_2=Z'_2VZ_1$ and the same identification (\ref{2j})
leads to the trivial relation $X_1X'_2=X'_2X_1$.

For the derivatives we may similarly introduce the realization
\begin{equation}\label{2k}
D =  \left( \begin{array}{cc}
          v^{-1} \partial_1 & u^{-1} \partial_2 \\
           u \partial_4 & v \partial_3
           \end{array} \right) \; , \quad \bar{D} =
\hat{V}^{\epsilon \, -1} D =  \left( \begin{array}{cc}
          v \partial_3 & -q^{-2} u^{-1} \partial_2 \\
          -q^{-2} u \partial_4 & v^{-1} \partial_1
           \end{array} \right) \; ,
\end{equation}
$$             \bar{ \partial } = \left( \begin{array}{cc}
          \partial_3 & -q^{-1} \partial_2 \\
          -q^{-1}  \partial_4 &  \partial_1
           \end{array} \right) \;,
$$
\noindent
(cf. (\ref{2j})) and find $ \bar{ \partial}_1 \bar{ \partial}_2 =
\bar{ \partial}_2 \bar{ \partial}_1 $.
Thus, the matrices $X$ and $\bar{ \partial}$ correspond to the
ordinary (commuting) coordinates and derivatives. Indeed,
it is not difficult to show that the mixed equation
\begin{equation}\label{2n}
\bar{D}_1VZ_2=Z_2 V \bar{D}_1 + 2 P_- \quad , \quad P_- = \frac{1}{2}
\left( \begin{array}{crrc}
        0&0&0&0 \\
        0&1&-1&0 \\
        0&-1&1&0 \\
        0&0&0&0
       \end{array} \right)
\end{equation}
\noindent
is invariant, and that the above `undressing' of $Z$ and $ \bar{D}$
reduces it to
\begin{equation}\label{2o}
\bar{ \partial}_1X_2=X_2 \bar{ \partial }_1 + 2P_-
\end{equation}
\noindent
which is just a rewriting of $ \partial_ix^j = x^j \partial_i + \delta^{j}_i$.
This shows that, in spite of the initial non-commutativity of the
entries of $Z$
and $D$, the use of  two non-commuting
variables ($u,v$) that commute with $x$ and $ \partial$
 allows us to `undress' ${\cal M}^{(3)}_q$ and
${\cal D}^{(3)}_q$. This possibility differs from that used in the
second reference in \cite{CHA-DEM}, where the relation between
the commutative and the non-commutative coordinates is given in terms
of a `$q$-bein', hence requiring 16 additional non-commutative
quantities.

At the algebra level one may
use \cite{LRTN} only one operator $q^L$ ($L$ being an element of
the classical Lorentz algebra) but, unlike $u$ and $v$, it
does not commute with the coordinates. In fact, the twisting \cite{DRINFELD}
of a Hopf algebra $A$
does not change its algebra structure (the multiplication and hence commutation
rules), while the coproduct is changed by the similarity transformation
$ \Delta^F (\cdot)= F \Delta ( \cdot) F^{-1}$ where  $F$ is an element
of $A \otimes A$ which satisfies  certain  requirements \cite{DRINFELD,RE}.
To preserve the $*$-structure of the Hopf algebra $A$
the element $F$ must be unitary,
$F^{\dagger}=F^{-1}$. It was shown \cite{LRTN} that the ${\cal M}_q^{(3)}$ case
is related to
the classical Poincar\'e algebra through  twisting.
In order to respect the $*$-structure {\it i.e.}, the reality condition,
an $F$ different  from that  in  \cite{LRTN} must be used.
Let us find this $F$ explicitly. The commutators of the Poincar\'e algebra
are given  by
\begin{equation}\label{tw1}
\begin{array}{l}
\, [L^{mn},L^{pk}]= g^{mk}L^{np}+g^{np}L^{mk}
                        -g^{mp}L^{nk}-g^{nk}L^{mp} \quad , \\
\, \\
\, [ \partial^k ,L^{mn}]=g^{km} \partial^n - g^{kn} \partial^m \quad , \qquad
[ \partial^k , \partial^l ]=0 \quad,
\end{array}
\end{equation}
\noindent
where the non-zero elements of the metric ($g_{ \mu \nu}=g^{ \mu \nu}$)
are $g^{13}=1=g^{31}$ and  $g^{24}=-1=g^{42}$ (see below eq.
(\ref{nmetric}) for $q=1$);
$\partial_1^{\dagger}=-\partial_1$, $\partial_3^{\dagger}=-\partial_3$,
$\partial_2^{\dagger}=-\partial_4$. With $L^{12}=-L_3$,
$L^{13}=L_5$, $L^{14}=-L_4$, $L^{23}=L_1$, $L^{24}=-L_6$ and $L^{34}=-L_2$
(to compare with \cite{LRTN}) we have
\begin{equation}\label{tw2}
L_1^{\dagger}=-L_2 \quad , \quad L_3^{\dagger}=-L_4 \quad , \quad
L_5^{\dagger}=-L_5 \quad, \quad L_6^{\dagger}=L_6 \quad .
\end{equation}
\noindent
All these Poincar\'e Lie algebra generators are `primitive elements' for the
coproduct {\it i.e.},
$ \Delta (X)=X \otimes 1+1 \otimes X$. To preserve the $*$-structure
of the Poincar\'e algebra in the twisting process  we introduce
\begin{equation}\label{tw3}
F=q^{\alpha \, L_5 \otimes L_6 } \quad , \quad  \alpha \in R \quad ,
\qquad F^{\dagger}=F^{-1} \quad .
\end{equation}
\noindent
This twisting operator, together  with the non-linear transformations
\begin{equation}\label{tw4}
L_i=(L_1,..., L_6) \mapsto (q^{\alpha L_6}L_1,L_2q^{\alpha L_6},
q^{- \alpha L_6}L_3,
L_4q^{- \alpha L_6}, L_5, L_6) \equiv \tilde{L}_i \quad,
\end{equation}
\noindent
\begin{equation}\label{tw5}
 \left( \begin{array}{cc}
            \partial_1 & \partial_2 \\
                   \partial_4  &  \partial_3
             \end{array} \right) \mapsto
\left( \begin{array}{cc}
            \partial_1 q^{\alpha L_6} & \partial_2 \\
             \partial_4 &  \partial_3 q^{- \alpha L_6}
             \end{array} \right) \equiv  \tilde{\partial} \quad ,
\end{equation}
\noindent
gives  (for $\alpha=-2$) the `$q$-deformed' Lorentz algebra and
the non-commuting traslations
   (the change in the form of the commutators is due to
the redefinitions (\ref{tw4}) and (\ref{tw5})) as well as  the coproduct
introduced in \cite{CHA-DEM} for the $\tilde{L}$'s and the
$\tilde{ \partial}$'s; it also preserves the hermiticity relations (\ref{tw2}).
We see here that the matrix $\tilde{ \partial}$ provides  another
possibility for the `dressing' of the $\partial$'s now in terms of one operator
(cf. (\ref{2k})), since its entries
satisfy the same commutation relations as those of $D$ (see (\ref{2h}))
and $\tilde{\partial}$ is also antihermitian.

\section{Physical considerations}

\indent
Let us now discuss some physical problems of the $q$-deformation of spacetime,
and specially of the  Minkowski space ${\cal M}_q^{(3)}$ as the simplest
example.
Due to its twisted character,  we have seen that there is a basis for
this deformation which coincides with the basis of the standard Poincar\'e
algebra. This might lead us to believe that there is nothing new in this case.
However, the global transformations in the deformed case introduce new
non-commutative quantities (operators) and this will result in new
features as it will be shown below.

Let us first discuss the quasiclassical limit {\it i.e.}, the transition from
the
commutation relations  to the corresponding Poisson
brackets; this limit is performed similarly in all  cases
${\cal M}^{(i)}_q$ from the appropriate eq. (\ref{2c}). The limit $q=e^{ \gamma
\hbar} \rightarrow 1$, $\hbar \rightarrow 0$,
is governed by the correspondence  $1/i \hbar \, [\hat{A}, \hat{B}]
\rightarrow \{ A,B \} $ from the operators $\hat{A}$, $\hat{B}$ to the
commuting quantities $A$, $B$ provided that the $R$-matrices
are normalized as
\begin{equation}\label{3a}
R^{(i)}= 1 + \gamma \hbar r^{(i)} + {\cal O}(\hbar^2) \quad,
\end{equation}
\noindent
$r^{(i)}$ being their quasiclassical counterparts. Two novel features appear
as a result of the quasiclassical limit, irrespective of the specific
$q$-deformed
spacetime ${\cal M}_q^{(i)}$ considered. First, the Poisson brackets (PB) of
$x^i$ and
$p_i$ (given by the limit of the commutation rules established by the RE for
coordinates and  derivatives) which now are $c$-numbers, become non-canonical,
the departure from the canonical ones being governed by the new constant $
\gamma$.
Secondly, although $K'=MKM^{\dagger}$ becomes the usual Lorentz transformation
of spacetime
(for $K= \sigma_{\mu}x^{\mu}$), the elements of the usual
Lorentz group now have non-trivial PB in general. For instance, it follows
from (\ref{2d})
\begin{equation}\label{3b}
\, \{ M_1,M_2^{\dagger} \} = i \gamma ( M_1r^{(2)}M_2^{\dagger}-
M_2^{\dagger}r^{(2)}M_1) \quad.
\end{equation}
\noindent
For the case of ${\cal M}_q^{(3)}$, to which we shall restrict from now on,
 $r^{(1)}=r^{(4)}=0$ and
$r^{(2)}=r^{(3)}=r=diag(1,0,0,1)$.

The PB for coordinates $Z$ and momenta $P$  (which transforms as the
derivative matrix $D$) for ${\cal M}_q^{(3)}$ are  obtained
from the quasiclassical limits of
(\ref{2h}) where $D$ is replaced by $P/(-i \hbar)$
and $\delta_i$ by $p_i/(-i \hbar)$
(cf. (\ref{2h1}))  ($P=P^{\dagger}$). In this way, for   $q=1$
(\ref{2h}) reduces to
$[ \hat{x}^i, \hat{p}_j]=i \hbar \delta^i_j$.   The quasiclassical limits now
give
\begin{equation}\label{3c}
\begin{array}{l}
\{ Z_1,Z_2 \} = i \gamma [Z_1rZ_1, {\cal P}] \\
\{ P_1,P_2 \} = -i \gamma [P_1rP_1, {\cal P}] \\
\{ P_1,Z_2 \} = i \gamma [P_1Z_2 ,r] - {\cal P} \quad.
\end{array}
\end{equation}
\noindent
These PB are invariant under the usual Lorentz transformations provided  that
the entries of $M$ and  $M^{\dagger}$ have zero Poisson
brackets with  those of $Z$ and $P$  (which has to be the case
since they already commuted in the $q$-case)
and (\ref{3b}) is valid. But as eqs. (\ref{3b}) and (\ref{3c}) (and similar
ones for the  other $q$-spacetimes ${\cal M}_q^{(i)}$) show, even at the
classical level a deformed
Minkowski phase space cannot be reduced to the classical one (as a
dynamical or  invariant relativistic system). In particular, the
parameters of the Lorentz group have non-trivial Poisson brackets
(\ref{3b}) among themselves; we  obtain a Lie-Poisson group
\cite{DRINFELD,STS} rather than a Lie group and its homogeneous
Poisson space.

To discuss the Hamiltonian dynamics of a particle we shall assume  the Dirac
constraint formalism. We shall take the mass-shell constraint $\varphi=
{\bf{p}}^2- m^2=(p_1p_3-q^{-2}p_2p_4)-m^2$ as usual, and introduce the
additional
one $\varphi'= (z^1- \tau)$ to eliminate  the unwanted degree of freedom
and separate an evolution parameter. This means that we use
light-cone-like variables as suggested by the real elements of $Z$;
we shall take $p_1= (m^2 + |p_2|^2)/p_3$ as energy in the quasiclassical
limit. In this picture,
the 2$\times$2 PB matrix $C$ of the constraints is specially simple,
$C= \{ \varphi , \varphi' \} (i \sigma_2)=-ip_3 \sigma_2$. The new (Dirac)
PBs are obtained from
\begin{equation}\label{3d}
\{ A,B \}^* = \{ A,B \} -( \{ A,  \varphi \} \,,\, \{ A,\varphi' \}) C^{-1}
( \{  \varphi ,B \} \,,\, \{ \varphi' ,B \} )^t \quad.
\end{equation}
\noindent
Since $ \{ P, \varphi \} =0$ (the d'Alembertian was already central), the
PB of momenta do not change. Then, from the standard Hamiltonian equations
$ \dot{A}= \{ H,A \} $ we obtain with  $p_1$ as Hamiltonian
\begin{equation}\label{3e}
\left( \begin{array}{cc}
                         0 & v^4 \\
                         v^2 & v^3
                         \end{array} \right) = \{ p_1,Z \}^*=
\left( \begin{array}{cc}
                         0 & -p_2/p_3 \, - i \gamma p_1z^4 \\
                     -p_4/p_3 \, + i \gamma p_1z^2    & p_1/p_3
                         \end{array} \right) \quad.
\end{equation}
\noindent
Thus, even in
this simple case, we obtain rather surprising expressions for the
velocities $v^{2,4}$ which for $\gamma \neq 0$  depend on the coordinates.
For the momenta we get
\begin{equation}\label{3f}
\dot{P} = \left( \begin{array}{cc}
                         \dot{p}_1 & \dot{p}_2 \\
                         \dot{p}_4 & \dot{p}_3
                         \end{array} \right) =
\{ p_1,P \}= i \gamma \left( \begin{array}{cc}
                         0 & -p_1p_2 \\
                         p_1p_4 & 0
                         \end{array} \right) \quad.
\end{equation}
\noindent
Hence,
\begin{equation}\label{3g}
\begin{array}{l}
p_1( \tau) = \mbox{const} \quad , \quad p_3( \tau) = \mbox{const} \quad , \\
\, \\
p_2( \tau)= \exp (-i \gamma p_1 \tau ) p_2(0)=(p_4( \tau))^*
\end{array}
\end{equation}
\noindent
($p_2=(p_4)^*$ is not conserved) and
\begin{equation}\label{3g1}
\begin{array}{l}
z^3(\tau)= {\displaystyle \frac{p_1}{p_3}} \tau + z^3(0) \quad ,\\
\, \\
z^2(\tau)=(z^2(0)- {\displaystyle \frac{ p_4(0)}{ p_3}} \tau)
\exp (i \gamma p_1 \tau ) = (z^4( \tau))^* \quad.
\end{array}
\end{equation}

Although the limit  $ \gamma =0$ reproduces the standard constant momenta
and linear evolution of coordinates with respect to the parameter $\tau$,
the $\gamma$-deformed behaviour is, even in this simple case,
strongly oscillating in the ($x,y$) plane and may relate points separated by
spacelike intervals. Hence the dynamical trajectories
do not coincide with the trajectories of the Lie group action (straight
lines for translations). This is because the group action vector fields
are  not Hamiltonian ones due to the Lie-Poisson group nature of the problem.
We stress that the PB for the other $q$-spacetimes  are similar to
(\ref{3b}), (\ref{3c}) and that our treatment is general. However,
the equations of the motion are less transparent than those for
${\cal M}^{(3)}_q$.
It is worth recalling  that the crucial point in the previous analysis
was the assumption $q \simeq 1 + \gamma \hbar$ between the deformation and
Planck constants; in the absence of a definite `correspondence principle'
intertwining deformation and quantization, other assumptions could be possible.
For instance, if the
dependence of $\hbar$ is of higher order, $(q-1)/ \gamma \hbar
\rightarrow 0$ for $\hbar \rightarrow 0$, no trace of the deformation
survives in the classical theory. If, on the other hand, $q$ does
not depend on $\hbar$ ({\it i.e.}, the deformation is completely unrelated
to quantization), then one needs an analogue of the classical mechanics for
non-commuting $q$-numbers.

To complement the description of the dynamics for the simple ${\cal M}_q^{(3)}$
case, we now look at the first step towards a field theory, the free wave
equation (for a discussion of $q$-wave equations see \cite{PW,MW}).
In our framework, such an equation must translate the constraint $\varphi$ into
a condition on the
wavefunctions $ \Phi (Z)$,
\begin{equation}\label{3h}
( \Box_q + m^2) \Phi (z^1,...,z^4)=0 \quad,
\end{equation}
\noindent
where $ \Box_q$ is the deformed d'Alembertian, $ \Box_q = 1/2 \,tr( \bar{D}D)=
det_qD$. Let {\it now}  $\bar{P}$ be the contravariant
2$\times $2 matrix of the {\it eigenvalues}
of $\bar{D}$, $\bar{P} \mapsto M\bar{P}M^{\dagger}$, with commuting
properties given by
\begin{equation}\label{3i}
\bar{P}_1V\bar{P}_2=\bar{P}_2V\bar{P}_1 \quad, \quad
\bar{P}_1VZ_2=Z_2V\bar{P}_1 \quad, \quad D_1\bar{P}_2=V\bar{P}_2D_1V^{-1}
\end{equation}
\noindent
({\it i.e.}, the same as those for $Z$ and $\bar{D}$ but without
the inhomogeneous
term in the third eq. in (\ref{2h})). The scalar product ($P,Z$)=($Z,P$)
between
the momenta $P$ and the coordinates $Z$ is given (cf. (\ref{2m})) by
\begin{equation}\label{3j}
(P,Z) \equiv \frac{1}{2} tr(PZ)=\frac{1}{2} tr( \bar{Z} \bar{P})=
\frac{1}{2}(p^1z^3+p^3z^1-q^2(p^2z^4+p^4z^2))\;,
\end{equation}
\noindent
where again the covariant $P$ is given by $P \equiv \hat{V}^{\epsilon}\bar{P}$:
\begin{equation}\label{3k}
\bar{P}= \left( \begin{array}{cc}
            p^1 & p^4 \\
            p^2 & p^3
            \end{array} \right) \quad , \qquad
P = \hat{V}^{\epsilon} \bar{P} = \left( \begin{array}{cc}
            p^3 & -q^2p^4 \\
            -q^2p^2 & p^1
            \end{array} \right) \quad .
\end{equation}
\noindent
The scalar product (\ref{3j}) is invariant and central,
$[(P,Z),Z]=0=[(P,Z),P]$.
Since
\begin{equation}\label{3l}
\Box_q (P,Z)^n= (P,Z)^n \Box_q + 2n(P,Z)^{n-1}(P,D) + n(n-1)(P,Z)^{n-2}(P,P)
\quad,
\end{equation}
\noindent
acting on the unity at the right ($D.1=0$) this implies that
\begin{equation}\label{3m}
\Box_q \sum_n \frac{(iP,Z)^n}{n!}= - \sum_n \frac{(iP,Z)^{n-2}(P,P)}{(n-2)!}
\quad.
\end{equation}
\noindent
This means that we may also define here the equivalent
of the Klein-Gordon plane wave since
\begin{equation}\label{3n}
( \Box_q + m^2) \exp \,i(P,Z)=(m^2-(P,P)) \exp \, i(P,Z) =0
\end{equation}
\noindent
when the mass-shell constraint $(P,P)=m^2$ is fulfilled.
It is interesting to point out that the commutation relations of $Z$ and $D$
with the dilatation operator (\ref{2f}) for ${\cal M}_q^{(3)}$, $s=tr(ZD)$,
are the same as the undeformed ones ($sZ=Z(s+1)$, $sD=D(s-1)$)
whereas for ${\cal M}_q^{(1,2)}$ they include the deformation parameter
$q$ \cite{OSWZ,AKR2}.

The fact that the non-commuting factors $u$, $v$ drop from the scalar
(invariant) products and from the product of the four $dp^i$ allows us to write
\begin{equation}\label{3o}
\Phi_q(Z) = \int d^4{\bf{p}} \, \delta ( {\bf{p}}^2
-m^2)[a_q(p)e^{i(P,Z)}
+ h.c. ] \quad ,
\end{equation}
\noindent
where $d^4 {\bf{p}}$  and $\delta ( {\bf{p}}^2 -m^2)$ are  the ordinary
integral measure
and mass shell delta function respectively.
This $q$-scalar field thus depends in practice on {\it commuting}
coordinates and momenta  due
to the cancellation of the ($u$, $v$) factors in the
`undressing' process; its only difference
with respect to the standard Klein-Gordon field is in the presence of the
twisting
parameter $q$ in the scalar products. This produces a trivial deformation
of the pole structure of the na\"{\i}ve Green functions of the theory
associated with the
re-definitions $p^{2,4} \rightarrow qp^{2,4}$ of the transverse momenta and
coordinates but, apart from this, it looks like the ordinary free theory.
However, a closer inspection of this and the other ${\cal M}_q^{(i)}$
cases reveals that a proper definition of the Green functions is
lacking. Moreover, due to the peculiarities of the quasiclassical limit,
a possible path integral derivation of the Green functions appears
to be fraught with    great difficulties related to the canonical
measure  and action functional.

\section{Conclusions}

We have considered in an unified way several deformed spacetime algebras
${\cal M}^{(i)}_q$, the different possibilities being related to the
two $R$-matrices in (\ref{2d}). Although ${\cal M}^{(3)}_q$,
for instance, is very close to the standard case, and ${\cal P}^{(3)}_q$
has the same irreducible unitary representations, on the whole the
physical picture  is different from the usual one. The assumption of
the standard Dirac bracket formalism and the corresponding Hamiltonian
formulation gives rise to quadratic Poisson brackets  of coordinates
and momenta reflecting the Lie-Poisson nature of the situation. This leads
to  trajectories which coincide with the classical ones only when
the new parameter $\gamma$ is set to zero; similar features appear in other
cases with a more complicated $R$-matrix structure. In fact, the situation for
the ${\cal M}_q^{(1,2)}$ cases is worse: for instance,
it is not possible to define {\it simultaneously}
${\cal M}_q^{(1,2)}$ and ${\cal D}_q^{(1,2)}$ with the usual hermiticity
properties under the star operation \cite{OSWZ,AKR,AKR2}, which leads to
rather unusual momenta (or coordinates).
As for the solutions
of the  $q$-deformed Klein-Gordon equation as  given by the $q$-d'Alembertian
operator, the wave expansion requires the introduction of $q$-number parameters
and the corresponding $q$-integration. This is not a problem for the
simple ${\cal M}^{(3)}_q$ case  where $q$-plane waves may be easily
constructed,
but the situation  is much  more complicated for other
$q$-spacetimes.

In fact, it is not clear what
should be the specific physical criteria
(beyond the general ones in Sec.1) that would select the appropriate
physical $q$-spacetime.
As already mentioned, different $q$-Lorentz groups (see \cite{WOZA}) exist,
and a {\it mathematical} classification of quantum Poincar\'e groups
and their corresponding $q$-Minkowski spaces has been recently given
in \cite{POWOR}. Although this classification includes the example
\cite{CHA-DEM} specially discussed here and others such as the
$\kappa$-Poincar\'e group \cite{LRTN2}, it does not incorporate the case of
\cite{OSWZ} although, for instance, both the $q$-Minkowski
spaces of \cite{OSWZ} and \cite{CHA-DEM} appear as special ones in
our framework.
Moreover, a detailed analysis of the {\it physical} contents of the different
deformed spacetimes is lacking despite the fact that they may  give rise
to unusual properties  such as non-commutative time (see \cite{MARU}
for the case of $\kappa$-Minkowski space). Our discussion indicates that a
solution to these problems requires a better
understanding of the possible relation  between
$q$-deformation and quantization,  and   that more work is needed
to investigate whether a
$q$-deformed field theory  based in this approach is feasible.

\vspace{1\baselineskip}

\noindent
{\bf Acknowledgements:} This research has been partially
sponsored by a CICYT (Spain) research grant. One of us (P.P.K.)
also wishes to thank the DGICYT, Spain, for financial support.


\begin{thebibliography}{99}

\bibitem{DFR} S. Doplicher, K. Fredenhagen and J. Roberts, {\it Spacetime
quantization induced by classical gravity}, preprint DESY 94-065 (1994)

\bibitem{WSSW} U. Carow-Watamura, M. Schlieker, M. Scholl and S. Watamura, Z.
Phys. {\bf C48}, 159 (1990); Int. J. Mod. Phys. {\bf A6}, 3081 (1991)

\bibitem{OSWZ} W. Schmidke, J. Wess and B. Zumino, Z. Phys. {\bf C52}, 471
(1991); O. Ogievetsky, W. B. Schmidke, J. Wess and B. Zumino, Commun.
Math. Phys. {\bf 150}, 495 (1992)

\bibitem{AKR} J.A. de Azc\'arraga, P.P. Kulish and F. Rodenas,
Lett. Math. Phys. {\bf 32}, 173 (1994)

\bibitem{FRT} L.D. Faddeev, N. Yu. Reshetikhin and L.A. Takhtajan, Alg. i Anal.
{\bf 1}, 178 (1989) (Leningrad Math. J. {\bf 1}, 193 (1990))

\bibitem{K-SK} P.P. Kulish and E.K. Sklyanin, J. Phys. {\bf A25}, 5963 (1992)

\bibitem{K-SA} P.P. Kulish and R. Sasaki, Progr. Theor. Phys. {\bf 89}, 741
(1993)

\bibitem{MBRA} S. Majid, J. Math. Phys. {\bf 34}, 1176 (1993); cf. ibid.
{\bf 32}, 3246 (1991)

\bibitem{MAJID} S. Majid, J. Math. Phys. {\bf 34}, 2045 (1993)

\bibitem{MEYER} U. Meyer, {\it The $q$-Lorentz group and braided coaddition
in $q$-Minkowski space}, DAMTP 93-45 (revised Jan. 94); S. Majid and
U. Meyer, Z. Phys. {\bf C36}, 357 (1994)

\bibitem{LRTN2} J. Lukierski, H. Ruegg, V.N. Tolstoy and A. Nowicki,
Phys. Lett.  {\bf B264}, 331 (1991); see also J. Lukierksi, H. Ruegg and V.
Tolstoy, {\it Quantum $\kappa$-Poincar\'e 1994}, IC/94/250

\bibitem{K} P. P. Kulish, Algebra Anal. {\bf 6}, 195 (1994)

\bibitem{PSW} M. Pillin, W. Schmidke and  J. Wess, Nucl. Phys. {\bf B403}, 223
(1993)

\bibitem{LRR} J. Lukierski, H. Ruegg and W. R\"uhl, Phys. Let. {\bf B313},
357 (1993)

\bibitem{BACRY} H. Bacry, Phys. Lett, {\bf B306}, 41 (1993); ibid. {\bf 317},
523 (1993)

\bibitem{BMT} L.C. Biedenharn, B. Mueller and M. Tarlini,
Phys. Lett. {\bf B318}, 613 (1993)

\bibitem{AKR2} J.A. de Azc\'arraga, P.P. Kulish and F. Rodenas,
{\it Quantum groups and deformed special relativity}, FTUV-94-21/IFIC-94-19
(April 1994, hep-th/9405161)

\bibitem{BFC} L. Baulieu and E. Floratos, Phys. Lett. {\bf B258}, 171 (1991);
L. Castellani, Phys. Lett {\bf B327}, 22 (1994)

\bibitem{MS} G. Mack and V. Schomerus, Nucl. Phys. {\bf B370}, 185 (1992)

\bibitem{RS} N. Yu. Reshetikhin and F. Smirnov, Commun. Math. Phys. {\bf 131},
157 (1990)

\bibitem{MARU} S. Majid and H. Ruegg,  Phys. Lett. {\bf B334}, 348 (1994)

\bibitem{MAJ} S. Majid, {\it $q$-Euclidean space and quantum group Wick
rotation by twisting}, preprint DAMTP/94-03 (1994)

\bibitem{CHA-DEM} M. Chaichian and A. Demichev, Phys. Lett. {\bf B304},
220 (1993); J. Math. Phys. {\bf 36}, 398 (1995)

\bibitem{DRINFELD} V.G. Drinfel'd, in Proc. of the 1986 {\it Int. Congr. of
Mathematicians}, MSRI Berkeley, vol. {\bf I}, 798 (1987);
Leningrad Math. J. {\bf 1}, 1419 (1990)

\bibitem{ZUM} B. Zumino, {\it Introduction to the differential geometry of
quantum groups}, in {\it Mathematical Physics X}, K. Schm\"udgen ed.,
Springer-Verlag (1992), p. 20

\bibitem{LRTN} J. Lukierski, H. Ruegg, V.N. Tolstoy and A. Nowicki,
J. Phys.   {\bf A27}, 2389  (1994)

\bibitem{RE} N. Yu. Reshetikhin, Lett. Math. Phys. {\bf 20}, 331 (1990)

\bibitem{STS} M.A. Semenov-Tian-Shansky, Funct. Anal. Appl. {\bf 17}, 259
(1985); Publ. RIMS Kyoto Univ. {\bf 21}, 1237 (1985)

\bibitem{PW} M. Pillin, J. Math. Phys. {\bf 35}, 2804  (1994)

\bibitem{MW} U. Meyer, {\it Wave equations on $q$-Minkowski space},
DAMTP 94-10 (1994)

\bibitem{WOZA} S.L. Woronowicz and S. Zakrzewski,
Comp. Math. {\bf 90}, 211 (1994)

\bibitem{POWOR} P. Podl\'es and S.L. Woronowicz,  {\it On the classification
of quantum Poincar\'e groups}, hep-th/9412059, Dec. 1994

\end{thebibliography}
\end{document}